\documentstyle[epsfig,psfig,referee]{mn}
\def\nicefrac#1/#2{\leavevmode\kern.1em
\raise.5ex\hbox{\the\scriptfont0 #1}\kern-.1em
/\kern-.15em\lower.25ex\hbox{\the\scriptfont0 #2}}
\begin{document}
\title[Trispectrum of BOOMERanG Data]{The trispectrum of the Cosmic Microwave
Background on sub-degree angular scales: an analysis of the BOOMERanG data}

\author[G.~De Troia {\it et al.}]
{G.De Troia$^{1,7}$\thanks{\tt detroia@roma1.infn.it},
P.A.R.~Ade$^{2}$,
J.J.~Bock$^{3}$, J.R.~Bond$^{4}$,
A.~Boscaleri$^{5}$,\newauthor
C.R.~Contaldi$^{4}$, B.P.~Crill$^{6}$, P.~de
Bernardis$^{1}$, P.G. Ferreira$^{7}$,\newauthor
M.~Giacometti$^{1}$, E.~Hivon$^{8}$,
V.V.~Hristov$^{6}$, M. Kunz$^{7,9}$,
A.E.~Lange$^{6}$,\newauthor
S.~Masi$^{1}$,
P.D.~Mauskopf$^{2}$,
T.~Montroy$^{10}$,
P.~Natoli$^{11}$,
C.B.~Netterfield$^{12}$,\newauthor 
E.~Pascale$^{5}$,
F.~Piacentini$^{1}$, G.~Polenta$^{1}$,
G.~Romeo$^{13}$,
J.E.~Ruhl$^{10}$ \\
$^{1}$ Dipartimento di Fisica, Universit\`a La Sapienza, Piazzale
 A. Moro 2, I-00185 Roma, Italy \\
$^{2}$ Dept. of Physics and Astronomy, Cardiff University,
                Cardiff CF24 3YB, Wales, UK\\
$^{3}$ Jet Propulsion Laboratory, Pasadena, CA, USA \\
$^{4}$ Canadian Institute for Theoretical Astrophysics,
        University of Toronto, Canada \\
$^{5}$ IROE-CNR, Firenze, Italy \\
$^{6}$ California Institute of Technology, Pasadena, CA, USA \\
$^{7}$ Astrophysics, University of Oxford, Keble Road, Oxford OX1 3RH, UK \\
$^{8}$ IPAC, California Institute of Technology, Pasadena, CA, USA \\
$^{9}$ Astronomy Centre, University of Sussex, BN1 9QJ, Brighton, UK \\
$^{10}$ Dept. of Physics, Case Western Reserve Univ., Cleveland, OH, USA \\
$^{11}$ Dipartimento di Fisica, Universit\`a Tor Vergata, Via della Ricerca 
       Scientifica,1 I-00133 Roma, Italy \\
$^{12}$ Depts. of Physics and Astronomy, University of Toronto, Canada \\
$^{13}$ Istituto Nazionale di Geofisica, Roma,~Italy}

\maketitle

\begin{abstract}
The trispectrum of the cosmic microwave background can be used to
assess the level of non-Gaussianity on cosmological scales. It
probes the fourth order moment, as a function of angular scale, of
the probability distribution function of fluctuations and has been
shown to be sensitive to primordial non-gaussianity, secondary
anisotropies (such as the Ostriker-Vishniac effect) and systematic
effects (such as astrophysical foregrounds). In this paper we
develop a formalism for estimating the trispectrum from high
resolution sky maps which incorporates the impact of
finite sky coverage. This leads to a series of operations applied
to the data set to minimize the effects of contamination due to
the Gaussian component and correlations between estimates at
different scales. To illustrate the effect of the estimation
process, we apply our procedure to the BOOMERanG data set and show
that it is consistent with Gaussianity. This work presents the
first estimation of the CMB trispectrum on sub-degree scales.

 \end{abstract}

\begin{keywords}
  cosmic microwave background- statistics
\end{keywords}

%{\tt To be done:
%\begin{itemize}
%\item references
%\item appendix (partially inaccurate, relevance, completeness?)
%\item description of results (to check)
%\item conclusions (to check)
%\item formula breaks and figures for two-column
%\item check formulas
%\end{itemize}}

\section{Introduction}
\label{Introduction} The Cosmic Microwave Background (CMB) has
become the observational tool of excellence for probing the
statistical nature of inhomogeneities in the universe. The small
deviations from homogeneity %a uniform Planck spectrum 
which have been detected
by over two dozen different experiments can be directly related to
the primordial origin of perturbations in the early universe and
therefore to fundamental physics at very high energies. A new
threshold was crossed in the experimental forum with the high
resolution, high sensitivity mapping of significant fractions of
the CMB sky by the BOOMERanG (de Bernardis et al. 2000) and MAXIMA
(Hanany et al. 2000) experiments.  
A careful analysis of the variance of
fluctuations in these maps has led to accurate estimates of the
angular power spectrum, far surpassing previous experimental
analyses on small angular scales. More recent results from
BOOMERanG (Netterfield et al. 2002, Ruhl et al. 2002),
MAXIMA (Lee et al. 2001),
and from other experiments like DASI 
(Halverson et al. 2002), CBI (Pearson et al. 2002), VSA (Grainge et al. 
2002), ACBAR (Kuo et al. 2002) and ARCHEOPS (Benoit et al. 2002) 
have posed our knowledge of the CMB angular power spectrum on even more
solid ground. However, there is more information in the CMB fluctuations
than what is provided by its power spectrum alone. The standard way to 
extract this information is to analyse high signal to noise maps of the
CMB field like those already produced by BOOMERanG or the ones that the
MAP\footnote{http://map.gsfc.nasa.gov/} satellite is expected to provide 
shortly.  

There is, therefore, strong motivation to attempt a
more detailed study of the CMB sky; in principle one would like a
complete characterisation of the probability distribution function
of the fluctuations in the CMB with the hope that it might probe
more fundamental features of the origin of structure in the
universe. For example one relatively stringent test of whether the
origin of fluctuations is due to a standard, single-field
inflationary model is if the CMB is a realization of a nearly
Gaussian distribution \cite{fnl}, while other models like the
curvaton might lead to a much larger non-Gaussian contribution
\cite{curvaton,bartolo,bernardeau1,bernardeau2}. Many secondary anisotropies like the
Ostriker-Vishniac (OV) effect \cite{ov} and the Sunyaev-Zeldovich
effect \cite{sz} can introduce measurable non-Gaussianity while
foregrounds and systematics may contribute as well.

A standard method of parametrising an arbitrary probability distribution
function is in terms of all its higher order moments.
%One considers
%ensemble averages of products of stochastic field with itself and
%an arbitrary number of positions.
In the case of statistically homogeneous and isotropic fields it
is more convenient to consider moments %averages of products 
of the Fourier
transform of the field; these symmetries will impose a set of
selection rules which pick out the true degrees of freedom. The
past few years have seen initial attempts at constraining these
moments by measuring them with the available data. A series of
analyses of the bispectrum (the cubic moment) of the COBE data
have revealed the presence of a non-Gaussian systematic
\cite{fmg,bzg,komatsu1}. This has been confirmed with an analysis of the
trispectrum, the quartic moment \cite{cobetri,komatsu}. But no primordial
non-Gaussianity was detected. On smaller angular scales 
analyses of the QMAP and QMASK\cite{park} and MAXIMA \cite{maxng,mariobi} 
data have shown that their observations are consistent with
the assumption that the CMB anisotropies are the result of an
isotropic Gaussian random process. 
The analyses of these
data sets have also revealed that statistical methods may be
sensitive to the data processing pipeline. Moreover, different
technical issues must be confronted if one is considering finite
sky coverage as opposed to full sky coverage.

The trispectrum, which we consider in this study, probes a rather
different kind of non-Gaussianity than the bispectrum;
\cite{ngcomp} found it to be very sensitive to point sources and
\cite{pat_ov} showed it to be a powerful probe of the OV effect. It can
also be used to detect weak lensing in the CMB \cite{lensing}.
%\begin{verbatim}
In particular, as described in \cite{bernardeau3}, weak lensing
does not introduce a three-point correlation function, meaning that its expected
bispectrum is zero. The first non trivial higher order correlation function
to detect weak lensing effects is the trispectrum.
%\end{verbatim}
Using the trispectrum to test for non-Gaussianity in
high-resolution CMB maps complements therefore other analyses
using the bispectrum, like \cite{mariobi}. The purpose of this
paper is to improve, extend and apply the method for estimating
the trispectrum first proposed in Kunz {\it et al} (2001) where it
was applied to the full sky map produced by the COBE satellite.

Using techniques developed for performing the operation in pixel space
\cite{fms,sg99,hu01} we present a formalism which can
easily be extended to high $\ell$s and discuss a robust method
for identifying an orthogonal set of estimators in the full-sky case.
We then discuss the various
numerical and statistical problems one faces when analysing
finite sky coverage. We finally use the BOOMERanG data as a test case to
extract the first estimate of the trispectrum on sub-degree angular scales.
\section{Formalism and Notation}
\label{notation}
In this section we present the notation that will be used throughout this
work. We start with a temperature anisotropy field defined on the
sphere, $\frac{\Delta T}{T} ({\bf n})$; it may be expanded  in terms
of spherical harmonic functions, $Y_{\ell m}({\bf n})$:
\begin{eqnarray}
\frac{\Delta T}{T} ({\bf n}) = \sum_{\ell m} a_{\ell m} Y_{\ell m}({\bf n}) .
\label{eqn:eq1}
\end{eqnarray}
For any theory of structure formation, the $a_{\ell m}$ coefficients are
a set of random variables; we shall restrict ourselves to
theories which are statistically homogeneous and isotropic.
The power spectrum $C_\ell$ of the
temperature anisotropies is then defined by
\begin{eqnarray}
\langle a_{\ell m}
a^*_{\ell' m'}\rangle=C_{\ell}\delta_{\ell \ell'}\delta_{m m'}.
\end{eqnarray}
If we consider the 3-point function of the temperature field, we obtain
the bispectrum, defined as
\begin{eqnarray}
\langle a_{\ell_1 m_1}a_{\ell_2 m_2}a_{\ell_3 m_3}\rangle=
\left (\begin{array}{ccc}
    \ell_1 & \ell_2 & \ell_3 \\
     m_1   &  m_2   & m_3
       \end{array}
\right) B_{\ell_1\ell_2\ell_3} .
\end{eqnarray}
The term ($\cdots$) is a Wigner 3J symbol, which arises due to the ``selection rules''
of the moments.

Following the same steps, we can construct the 4-point function and the
associated trispectrum. We represent the rotationally invariant solution
for the trispectrum as in \cite{hu01}:
\begin{eqnarray}
\langle a_{\ell_1 m_1}a_{\ell_2 m_2}a_{\ell_3 m_3}a_{\ell_4 m_4}\rangle=
\sum_{LM}\left (\begin{array}{ccc}
    \ell_1 & \ell_2 &  L \\
     m_1   &  m_2   & -M
       \end{array}
\right)\left (\begin{array}{ccc}
    \ell_3 & \ell_4 &  L \\
     m_3   &  m_4   &  M
       \end{array}
\right)(-1)^M Q^{\ell_1 \ell_2}_{\ell_3 \ell_4}(L)
\label{eqn:eq4}
\end{eqnarray}
Using the orthogonality properties of the Wigner 3J symbols and the 
relation $Q=T+G$, we can
invert the equation (\ref{eqn:eq4}) to obtain the estimator
\begin{eqnarray}
\label{eqn:eq5}
\hat{T}^{\ell_1 \ell_2}_{\ell_3 \ell_4}(L)=(2L+1)\sum_{m_1 m_2 m_3 m_4 M}
(-1)^M \left (\begin{array}{ccc}
    \ell_1 & \ell_2 &  L \\
     m_1   &  m_2   &  M
       \end{array}
\right)\left (\begin{array}{ccc}
    \ell_3 & \ell_4 &  L \\
     m_3   &  m_4   &  -M
       \end{array}
\right) \times \\
a_{\ell_1 m_1}a_{\ell_2 m_2}a_{\ell_3 m_3}a_{\ell_4 m_4} -
\hat{G}_{l_3 l_4}^{l_1 l_2}(L) \nonumber
\end{eqnarray}

The term $G_{l_3 l_4}^{l_1 l_2}(L)$ represents the unconnected Gaussian contribution
and it is given in (Hu 2001) as:
\begin{eqnarray}
G_{l_3 l_4}^{l_1 l_2}(L)=(-1)^{(l_1+l_3)}\sqrt{(2l_1+1)(2l_3+1)}C_{l_1}C_{l_3}
\delta_{l_1l_2}\delta_{l_3l_4}\delta_{L0}+(2L+1)C_{l_1}C_{l_2}\times \\ \nonumber
[(-1)^{l_2+l_3+L}\delta_{l_1l_3}\delta_{l_2l_4}+\delta_{l_1l_4}\delta_{l_2l_3}] \nonumber
\label{eqn:gaus_term}
\end{eqnarray}

The term $\hat{T}^{\ell_1 \ell_2}_{\ell_3 \ell_4}(L)$ is the connected part
of the angular trispectrum and its expectation value is exactly zero for a
Gaussian field. This means that the connected part is sensitive to the presence
of non Gaussianities.
The unconnected term is non-zero only for $L=0$ or $l_1=l_2=l_3=l_4$,
but only with full sky coverage. In the case of incomplete sky coverage
the unconnected terms can contaminate all other modes.
We will discuss this situation in section (\ref{hires}).

%\begin{verbatim}
For the purpose of this work we have not computed the possible trispectrum
components. We concentrated only on the simpler case $\ell_1=
\ell_2=\ell_3=\ell_4=\ell$, i.e. the diagonal component.
Recent papers have investigated in detail the power of the diagonal trispectrum 
in the presence of some non-Gaussian signals mentioned in the introduction.
In particular, \cite{ngcomp} has shown that for simulated point-source maps
the diagonal trispectrum is much more powerful than the nearly diagonal estimator
($\ell,\ell+1,\ell+2,\ell+3$), even though the latter does not contain a Gaussian
contribution.
Also, in  \cite{pat_ov} it is discussed
how the OV effect generates a signature on the diagonal tispectrum
which could easily be detected on the arcminute scales probed by the 
{\sc Planck}\footnote{http://astro.estec.esa.nl/SA-general/Projects/Planck/} mission.

Finally it should be noticed that computing all components of the trispectrum 
is a serious computational challenge.
Many modes are also correlated due to the limited sky coverage. For these
reasons we have decided to restrict this analysis to the case of $\ell_1=
\ell_2=\ell_3=\ell_4=\ell$. 
%\end{verbatim}

We start with the method described in
\cite{cobetri}. We define
\begin{eqnarray}
\langle a_{\ell m_1}a_{\ell m_2}a_{\ell m_3}a_{\ell m_4}\rangle=
\sum_{a=0}^n T_{\ell;a}\tau_{m_1 m_2 m_3 m_4}^{a;\ell}
\end{eqnarray}
where the $T_{\ell,a}$ are the components of the trispectrum that we wish to estimate,
and $\tau$ is a tensor which we have to determine in order to construct an estimator for
$T_{\ell,a}$. The geometrical considerations stated above, together with
the required symmetries with respect to the interchange of $\{\ell_i,m_i\}$ pairs suggest
\begin{eqnarray}
\bar{\tau}_{m_1 m_2 m_3 m_4}^{\alpha,\ell}=\sum_{M=-2\alpha}^{2\alpha}
(-1)^M \left (\begin{array}{ccc}
    \ell & \ell &  2\alpha \\
     m_1   &  m_2   &  M
       \end{array}
\right)\left (\begin{array}{ccc}
    2\alpha & \ell &  \ell \\
    -M   &  m_3   &  m_4
       \end{array}
\right) + \mathrm{ineq. permut.}
\end{eqnarray}
Although the $\bar{\tau}$s satisfy all the correct symmetries,
they define an over-complete basis. To correct for this
deficiency we define an orthonormalised set of tensors
\begin{eqnarray}
\tau_{m_1 m_2 m_3 m_4}^{a;\ell}=\sum_{\alpha=0}^{\ell}{\cal L}_\ell^{a\alpha}
\bar{\tau}_{m_1 m_2 m_3 m_4}^{\alpha,\ell} ,
\label{eqn:eq7}
\end{eqnarray}
where the matrix ${\cal L}_\ell^{a\alpha}$ is derived from the required
property that the $\tau$ be orthogonal with respect to the product
given in equation (5) of Kunz et al (2001). The estimator of the
trispectrum is then given by
\begin{eqnarray}
\hat{T}_{\ell;a}&=&\sum_{m_1 m_2 m_3 m_4}{\tau}^{a;\ell}_{m_1 m_2 m_3 m_4}
a_{\ell m_1}a_{\ell m_2}a_{\ell m_3}a_{\ell m_4} \\ \nonumber
&=&\sum_{m_i}\sum_{\alpha=0}^\ell {\cal L}^{a;\alpha}_\ell
\bar{\tau}^{\alpha;\ell}_{m_1m_2m_3m_4}
a_{\ell m_1}a_{\ell m_2}a_{\ell m_3}a_{\ell m_4} \\ \nonumber
&\equiv&\sum_{\alpha=0}^\ell {\cal L}^{a;\alpha}_\ell\bar{T}^{\alpha;\ell}
\end{eqnarray}
Note that there are only ${\rm int}(\ell/3)$ independent estimators
due to the symmetry properties of the $a_{\ell m}$.
%In Kunz {\it et al} (2001), we used the explicit definition of
%Eq. \ref{eqn:eq7} and evaluated a the set of relevant Wigner 3J coefficients.
In this paper we will consider the ``normalised'' trispectrum used
in Kunz {\it et al} (2001) where we divide each estimate of the trispectrum
by $({\hat C}_\ell)^2$, where
${\hat C}_\ell=\frac{1}{2\ell+1}|a_{\ell m}|^2$. Its statistical properties
are equivalent to the ones of the unnormalised estimator, and
it is more robust with respect to fluctuations in the power spectrum
\cite{ngcomp}.

\section{Application to high resolution maps with incomplete sky coverage}
\label{hires} In this paper we will be focusing on a
high-resolution map with incomplete sky coverage, in particular
the BOOMERanG data set. This leads to a set of algorithmic
problems which did not have to be addressed in Kunz {\it et al}
(2001). The three problems we wish to highlight are:
\begin{description}
\item[{\bf Speed:}] The numerical evaluation of Wigner 3J coefficients for large values of
$\ell$ becomes time consuming and practically unfeasible. Indeed beyond the
COBE resolution of a maximum $\ell$ of approximately 25 it is not possible
to estimate the $\bar{T}^{\alpha;\ell}$ sufficiently rapidly for a robust
Monte Carlo assessment of the statistics.
\item[{\bf Gaussian Contamination:}] The finite sky coverage will induce correlations between the estimators
with different values of $\ell$ and $a$ (or $\alpha$).
As a consequence, all estimators may be heavily contaminated by the Gaussian
(or disconnected) contributions to the maps.
\item[{\bf Correlations:}] The correlations between modes in the cut sky mean that
the $\bar{T}^{\alpha;\ell}$ will be even more correlated than in the
full sky case.
\end{description}

We shall now focus on the solutions we propose to these three problems
\subsection{Speed}

We have opted to use the method described in \cite{hu01,sg99}
for calculating $\bar{T}^{\alpha;\ell}$:
we define a new set of sky maps weighted in rings centred
around a point $\hat{\bf q}$:
\begin{eqnarray}
e_{\ell}(\hat{\bf q})=\sqrt{\frac{2\ell+1}{4\pi}}\int{d\hat{\bf n}
T(\hat{\bf n})P_{\ell}(\hat{\bf n}\cdot\hat{\bf q}})
\label{eqn:eq10}
\end{eqnarray}
To implement this method we start with the relation (\ref{eqn:eq10})
and we use the relation (\ref{eqn:eq1}) to express the temperature
$T$ as a function of spherical harmonics and the relation:
\begin{eqnarray}
P_\ell(\hat{\bf n}\cdot\hat{\bf q})=\frac{4\pi}{2\ell+1}\sum_m
Y_{\ell m}^{\ast}(\hat{\bf n}) Y_{\ell m}(\hat{\bf q})
\end{eqnarray}
to express also the Legendre polynomials as a function of spherical harmonics.
Combining them with equation (\ref{eqn:eq10}) we obtain:
\begin{eqnarray}
e_\ell(\hat{\bf q})=\sqrt{\frac{4\pi}{2\ell+1}}\sum_m a_{\ell m}
Y_{\ell m}(\hat{\bf q})
\label{eqn:e_l}
\end{eqnarray}
The $e_\ell$ calculation is quite fast because we can use
the fast Fourier transform on rings of equal latitude (Muciaccia et al 1997).

We can then rewrite equation (\ref{eqn:eq5}) in terms of this new set of
sky maps (Komatsu 2002):
\begin{eqnarray}
\label{eqn:Tell}
\bar{T}^{\alpha;\ell}=\frac{1}{4\alpha+1}\sum_{M=
-2\alpha}^{2\alpha}t_{2\alpha,M}^
{\ell\ell\ast} t_{2\alpha,M}^{\ell\ell}
\end{eqnarray}
where
\begin{eqnarray}
t_{LM}^{\ell\ell}=\sqrt{\frac{2L+1}{4\pi}}
\left (\begin{array}{ccc}
    \ell & \ell &  L \\
     0   &  0   &  0
       \end{array}
\right)^{-1}\int{d\hat{\bf n}}{[e_{\ell}(\hat{\bf n})e_{\ell}(\hat{\bf n})]
Y^{\ast}_{LM}(\hat{\bf n})}
\end{eqnarray}
If we expand the Wigner 3J symbols in terms of spherical harmonics and
use the addition theorem we obtain:
\begin{eqnarray}
\bar{T}^{\alpha;l}=
N^{-1}_{\ell 2\alpha}\int{\frac{d^2\hat{n}}{4\pi}}\int{\frac{d^2\hat{q}}
{4\pi}e_{\ell}(\hat{n})e_{\ell}(\hat{n})e_{\ell}(\hat{q})e_{\ell}(\hat{q})
P_{2\alpha}(\hat{n}\cdot\hat{q})}
\end{eqnarray}
where
\begin{eqnarray}
N_{\ell L}=\frac{1}{3}
\left (\begin{array}{ccc}
    \ell & \ell &  L \\
     0   &  0   &  0
       \end{array}
\right)^2
\end{eqnarray}
We have thus computed a set of $\bar{T}^{\alpha;\ell}$ which we can
orthonormalise
to get the estimator for the trispectrum $\hat{T}^{a;\ell}$.
This method is very fast, especially when estimating the trispectrum at high
values of $\ell$. Note that direct evaluation of the Wigner 3J coefficients,
e.g. by recurrence relations, would result in an
${\cal O}(\ell^5)$ problem, requiring $\sim 10^{16}$ operations
for $\ell\sim 1000$.

\subsection{Gaussian Contamination}

Kunz {\it et al} (2001) found that the purely
Gaussian contribution to the trispectrum (the disconnected part)
corresponds to the $\alpha=0$ term. By orthogonalising all
other estimators with respect to this tensor it is possible
to remove the Gaussian contribution exactly on a map by map
basis. The resulting
estimators are only sensitive to non-Gaussian
contributions, i.e. in the case of Gaussian skies they would
have a zero expectation value. Additionally, the variance of this estimator
is shown to be minimal \cite{cobetri}.
In the cut sky case, there are strong cross correlations
between components of the trispectrum with different values of
$\ell$ and $\alpha$. In this case, the orthogonalisation method fails.
Given that the Gaussian contribution to
the trispectrum may be much larger than the non-Gaussian contribution,
it is essential that we remove it as completely as possible nonetheless.

To overcome these problems we have chosen to employ the following Monte Carlo
scheme: we generate an ensemble of maps
with the same angular power spectrum, sky coverage and noise
as the data maps we want to analyse. We estimate the $\bar{T}^{\alpha;\ell}$
from each map and calculate the mean of these quantities over
the whole ensemble. Let us denote this mean by  $\bar{T}^{\alpha;\ell}_G$.
We then use this quantity to correct for the Gaussian
contamination in the estimate of the trispectrum from the data by
defining $\bar{T}^{\alpha;\ell}_{GC}=\bar{T}^{\alpha;\ell}-
\bar{T}^{\alpha;\ell}_{G}$.
Note that, by so doing, we are removing the Gaussian contamination before
performing the full-sky orthogonalisation, i.e. before multiplying
by ${\cal L}$.

\subsection{Correlations}
The fact that there is only limited sky coverage also implies that
there will be correlations between values of the trispectrum
at different $\ell$s. This is shown in section \ref{results} below. 
To strongly suppress
the correlations and to end up with a simple covariance
matrix, we consider band averaged values of the trispectrum.
Since there is no a priori given band width, we study the cases
of $\Delta \ell=40$, $50$ and $60$ and $\Delta a=10$ and
$15$. The choice $\Delta\ell\simeq 50$ is consistent with previous
analyses of the BOOMERanG power spectrum (see, e.g., \cite{ruhl}).
In this way we can check the sensitivity
of the results to the chosen band size.

\section{Numerical Implementation and Consistency Tests}

The process we use is basically the same as in a number of
previous analyses \cite{kogut96,fmg,cobetri,mariobi}.
We generate an ensemble of Gaussian maps with the
same angular power spectrum and noise property as estimated from the 
BOOMERanG data and
the same sky coverage. We then apply
our estimators to the set of maps to obtain a distribution for
each estimator in the Gaussian case. In particular we characterise
the full distribution in terms of the mean values of the estimators
and the covariance matrix between them. These quantities are used
to define a standard multivariate $\chi^2$ as a goodness of fit.
The estimators of the trispectrum are then evaluated from the
BOOMERanG data. In section 5 we will discuss their behaviour.
The goodness of fit of these estimators is
compared against its distribution for
a {\it new} ensemble of Gaussian maps. From this comparison we
can quantify the confidence with which the data can be said
to be Gaussian from the point of view of our estimator.
It is clear that the numerical details of this process must be well
understood if we are to believe in our results. We
focus on the particularities of the analysis in this paper which
differ from previous analyses.

It is important to
compare the results using this hybrid pixel/harmonic analysis
with the standard methods which have been used before. We do
so by looking at the two lower order statistics, i.e.
we have calculated the power spectrum $C_\ell$ and the
bispectrum $B_{\ell\ell\ell}$ using this new approach
as well as summing up the 3J symbols.
The relevant expressions are:
\begin{eqnarray}
C_\ell=\frac{1}{4\pi}\int{d{\bf q}\left |e_{\ell}({\bf q})\right |^2}
\end{eqnarray}
and (Spergel et al. 1999)
\begin{eqnarray}
\left (\begin{array}{ccc}
    \ell & \ell &  \ell \\
     0   &  0   &  0
       \end{array}
\right) B_{\ell\ell\ell}=\int{d{\bf q}e_\ell({\bf q})e_\ell({\bf q})e_\ell({\bf q})}
\end{eqnarray}

\begin{figure}
\centerline{\epsfig{figure=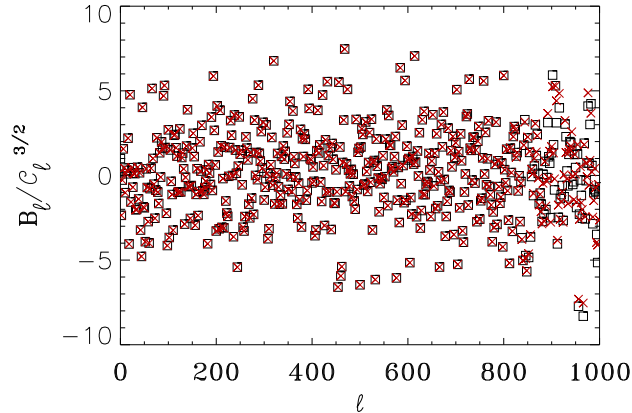,width=9cm}}
\caption{In
this figure we represent the normalised bispectrum
$I^3_\ell=B_\ell/C^{3/2}_\ell$ calculated for a full sky Gaussian
map with a pixel size of $7^{'}$. The red crosses show $I^3_\ell$
obtained with the $e_\ell$ method, the black boxes $I^3_\ell$ with
$a_{\ell m}$ and the 3J symbols. We can see that at $\ell\simeq
800$ there is a quite evident difference between two plots, due to
a pixelisation effect. We limit therefore our analysis to
$\ell\leq 700$.} \label{pix_eff}
\end{figure}
We have compared the  $C_\ell$ and $B_{\ell\ell\ell}$ using these
expressions with the standard results obtained
using $a_{\ell m}$ and the Wigner 3J symbols, for a maximum
$\ell$ value of $1000$.
Using a set of CMB Gaussian maps with the best fit power spectrum measured by Boomerang
\cite{netterfield} and
a pixel resolution of $7^{'}$ we have found that
the bispectrum
obtained with $e_\ell$ is affected by a pixelisation effect for high values
of $\ell$ (while the power spectrum shows no difference). 
To check for this,
we have done the same analysis with higher resolution ($\simeq 3^{'}$) and
have found that the pixelisation effect vanishes. Given that
we are restricted to the pixelisation level of the data,
we can use the comparison of the two estimates of the bispectrum
to define a maximum $\ell$ out to which we can trust the new
estimate of the trispectrum. Note that it is computationally intractable
to perform such a comparison in terms of the trispectrum, although
this would be preferable. From Fig.~\ref{pix_eff} one can see that discrepancies
arise for $\ell>800$ and we chose not to estimate the trispectrum
beyond $\ell=700$, leaving a conservative margin as we did not test
the trispectrum itself. Furthermore, we chose not to consider any $\ell$
below 100 because the BOOMERanG data are not very sensitive to these modes,
due to limited sky coverage and data filtering.

\begin{figure}%[h]%tb]
\centerline{\epsfig{figure=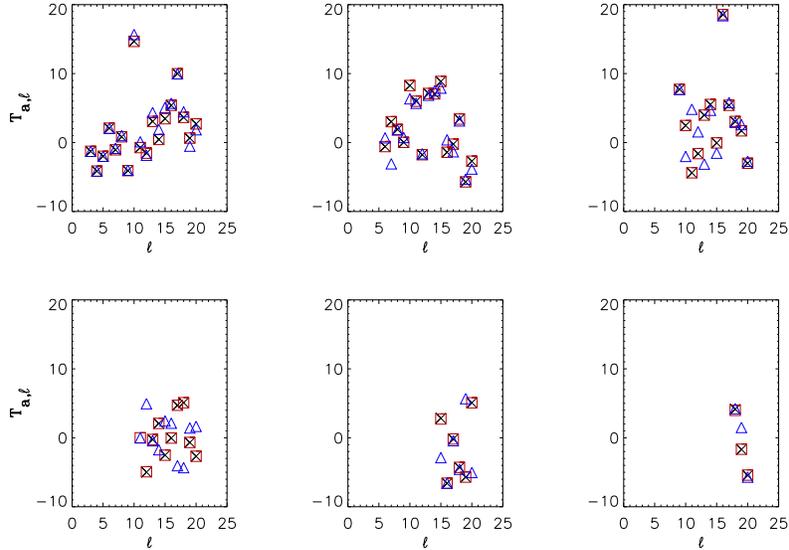,width=11cm}}
\caption{In this figure we reproduce the COBE results for the trispectrum with the
GS and SD orthonormalisation methods (see text) and compare them with (Kunz et al. 2001).The black
crosses are the (Kunz et al. 2001) results, green squares are GS method results and blue triangles
are SD method results.}
\label{cobe}
\end{figure}

Another novelty in our analysis (as compared to that of
Kunz et al. 2001) is the method for constructing
${\cal L}_{\ell}^{a;\alpha}$.
There, a Gram-Schmidt (GS) procedure was used to calculate the
orthonormal transformation matrix ${\cal L}_{\ell}^{a;\alpha}$.
Due to the inherent instability of the GS procedure,
it is not applicable to large matrices, i.e. for large $\ell$.
We have opted therefore to use an alternative orthonormalisation
method. We subtract the $a=0$ part and use a Jacobi routine to
obtain a spectral decomposition (SD) of the remaining matrix.
The eigenvectors of the non-vanishing eigenvalues form then
the transformation matrices. This method is robust
and, moreover, gives us an unambiguous procedure for
ordering the estimators through the different eigenvalues.
As a  strong consistency test we have applied our
trispectrum code to the COBE
data and compared the results with those of \cite{cobetri}.
The results of the SD method lie, up to a possible sign change,
very close to the original ones (see figure \ref{cobe}). In any case,
the statistical significance of the results (and the
conclusions one can extract) are the same as in Kunz et al (2001).
We advocate the use of the SD method from now on, even
in the case of analyses limited to low $\ell$s.

For our analysis we have used the best four of the six $150$~GHz
channels of the BOOMERanG 1998 flight; we naively coadd the data 
taken at the scan
speed of $1$ degree per second ($1$ dps). We simulate three sets
of 1000 Gaussian maps each. In fact, we need three statistically
independent ensembles of simulated maps for our analysis: one to
estimate the Gaussian contribution described above, another to
estimate our estimator's covariance matrix and a third one to
perform the actual comparison with real data. To produce the maps
we follow the very same steps used during the real Boomerang data
reduction.
%We make a map from the data by naively binning the
%timestream into
%pixels on the sky.
To generate a map we use timestream simulations
created with the actual flight pointing and transient flagging.
The signal component of these time streams is generated from
simulated gaussian CMB maps, while the noise component is from 
gaussian realizations
of the measured detector noise power spectrum. The noise spectrum is
determined using the iterative procedure described in Ferreira
$\&$ Jaffe (2000) and Netterfield et al.\ (2002).

To reduce the effects of $1/f$ noise on this
naively binned map, a brick-wall highpass Fourier filter is first
applied to the timestream at a frequency of $0.1$~Hz.
A notch filter is also applied between 8 and
9~Hz to eliminate a non-stationary spectral line in the data. This only
affects angular scales well above $\ell=1000$ and is therefore irrelevant
for our analysis.

The coaddition of several channels is achieved by averaging the maps
(both from the data, and from the Monte-Carlos of each channel).
Each channel has a slightly different beam size, which is
taken into account in the generation of the simulated maps.
We select the most central region of the scan
by applying an elliptical mask as in (Netterfield et al. 2002). This
corresponds to $\simeq 1.8\%$ of the full sky. The mask selects a region of
approximately uniform coverage and high signal to noise
and comprises $\sim 57000$ pixels of size $\sim 7'$ each in the Healpix
pixelization scheme (G\'{o}rski et al. 1998).
%The generation of a single
%realization can be achieved on the order of a minute on an IBM SP3.
We refer the reader to Ruhl et al.\ (2002)
for a thorough description of a simulation pipeline (based on the 
MASTER/FASTER algorithms described in \cite{hivon})
very similar to
the one used here.

\section{Results and application to the BOOMERanG data}
\label{results}

\begin{figure}
\vskip-1.3in
\centerline{\epsfig{figure=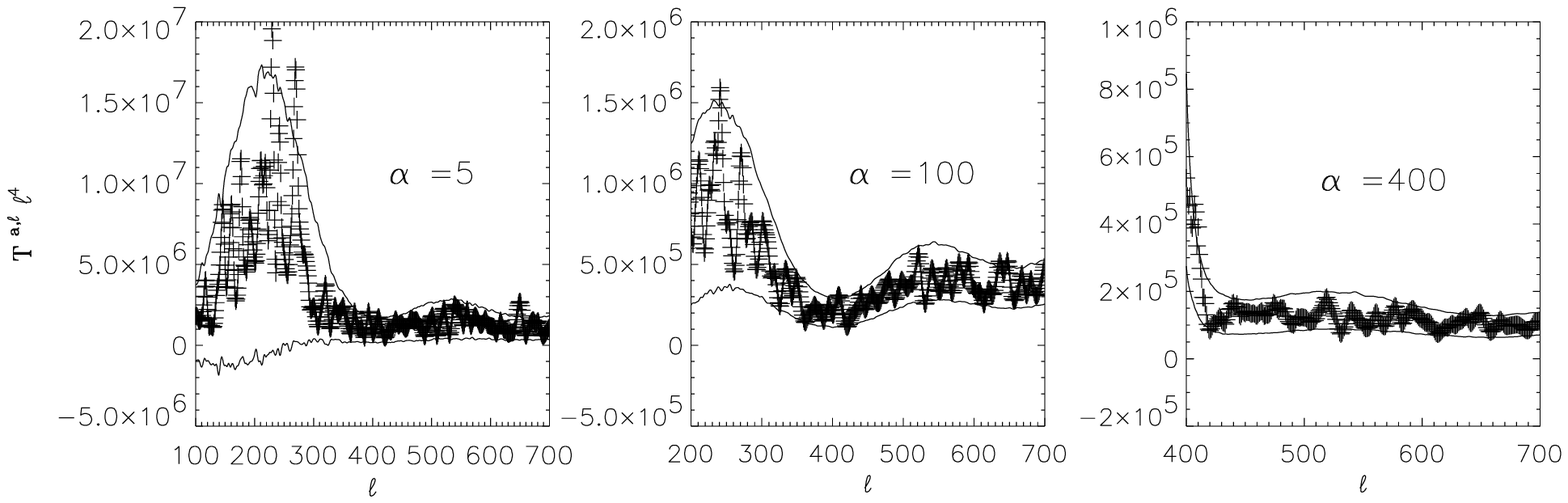,width=12cm}}
\vskip-1.4in
\centerline{\epsfig{figure=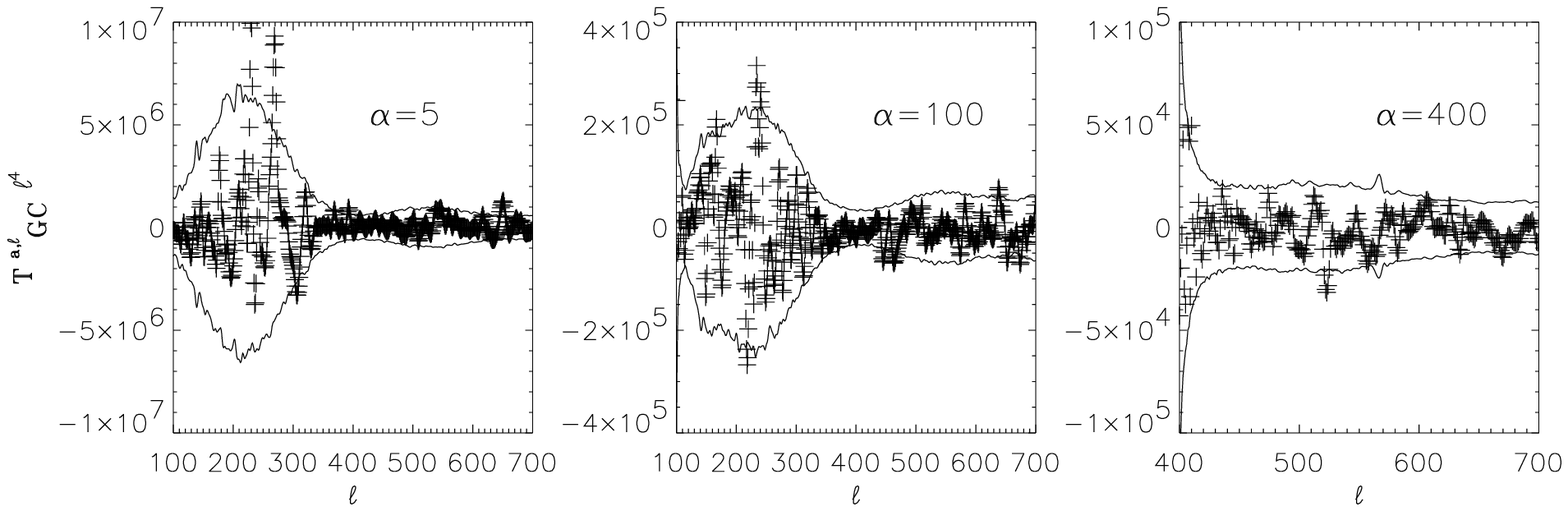,width=12cm}}
\caption{Top panel: An estimate of the non-orthogonalised
trispectrum, ${\bar T}^{\alpha;\ell}$ (multiplied by $\ell^4$) for $\alpha=5,100,400$ from
the BOOMERanG data (crosses) and the corresponding $95\%$
confidence limits from the $1000$ Monte-Carlo simulations. Bottom
panel: An estimate of the non-orthogonalised trispectrum corrected
for Gaussian contamination, ${\bar T}^{\alpha;\ell}$ (multiplied by $\ell^4$)
for
$\alpha=5,100,400$ from the BOOMERanG data (crosses) and the
corresponding $95\%$ confidence limits from the $1000$ Monte-Carlo
simulations. } \label{res1}
\end{figure}

To show in detail the method 
proposed in Sections 2 and 3 we are going to discuss the results
obtained at each step from both the data and the Monte-Carlo
simulations. We start with the estimate of the ${\bar
T}^{\alpha;\ell}$
 without
Gaussian corrections. In the top panel of Figure \ref{res1} we plot ${\bar
T}^{\alpha;\ell}$ as a function of $\ell$ for selected values of $\alpha$.
We can highlight two features. Firstly
the ${\bar T}^{\alpha;\ell}$ are highly correlated for adjacent
values of $\ell$ due to the finite sky coverage, as expected. Secondly,
and because of the finite sky coverage, there is a strong contamination
from the disconnected component of the trispectrum. This is evident in
the fact that the values of ${\bar T}^{\alpha;\ell}$ scatter
about the $(C_\ell)^2$ and that the $95\%$ confidence limits are not
centred about zero. As one would expect the lower the value of
$\alpha$, the more contaminated the estimate is by the disconnected part.
As advocated in Section \ref{hires}, we correct for the contamination
due to the disconnected component by using a Monte Carlo ensemble
(of 1000 realizations) to generate a correction. This can be seen
as a bias which must be subtracted off all estimates of
${\bar T}^{\alpha;\ell}$ with $\alpha>0$. In the bottom panel of
Figure \ref{res1}
we plot the ``Gaussian corrected'' estimate of
${\bar T}^{\alpha;\ell}$ with corresponding $95\%$ confidence limits.
As expected the estimates now scatter around zero, while the confidence
limits, although not necessarily symmetric around the $\ell$ axis are
much more centred. The remaining asymmetry is merely a manifestation that
for low $\alpha$s the distribution of the ${\bar T}^{\alpha;\ell}$ is
slightly skewed.

\begin{figure}
\vskip-1.3in
\centerline{\epsfig{figure=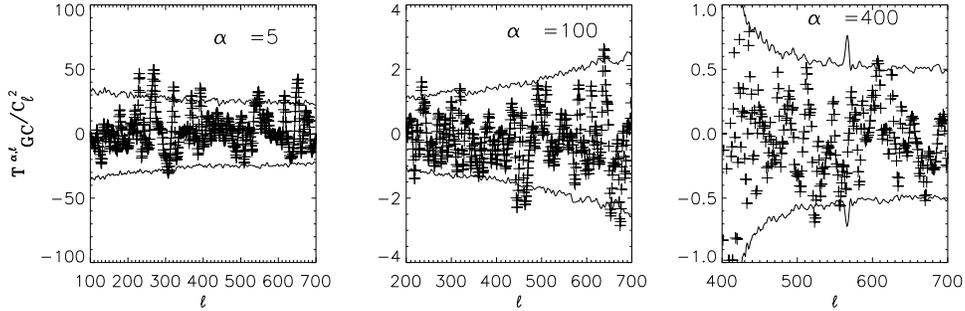,width=12cm}}
\caption{An estimate of the non-orthogonalised trispectrum
corrected for Gaussian contamination and normalised ${\bar
T}^{\alpha;\ell}/({\hat C}_\ell)^2$ for $\alpha=5,100,400$ from
the BOOMERanG data (crosses) and the corresponding $95\%$
confidence limits from the $1000$ Monte-Carlo simulations. }
\label{res2}
\end{figure}
Given that we will be working with normalised estimates of the
trispectrum (as in Kunz {\it et al} 2001) it is illustrative to
plot the $\ell$ dependence of ${\bar T}^{\alpha;\ell}/({\hat
C}_\ell)^2$ for a few values of $\alpha$. We do this in Figure
\ref{res2}. One can see the dependence on $\alpha$ of the $95\%$
confidence intervals, i.e. the $\ell$ value of minimum scatter
depends on $\alpha$.

\begin{figure}
\vskip-1.3in
\centerline{\epsfig{figure=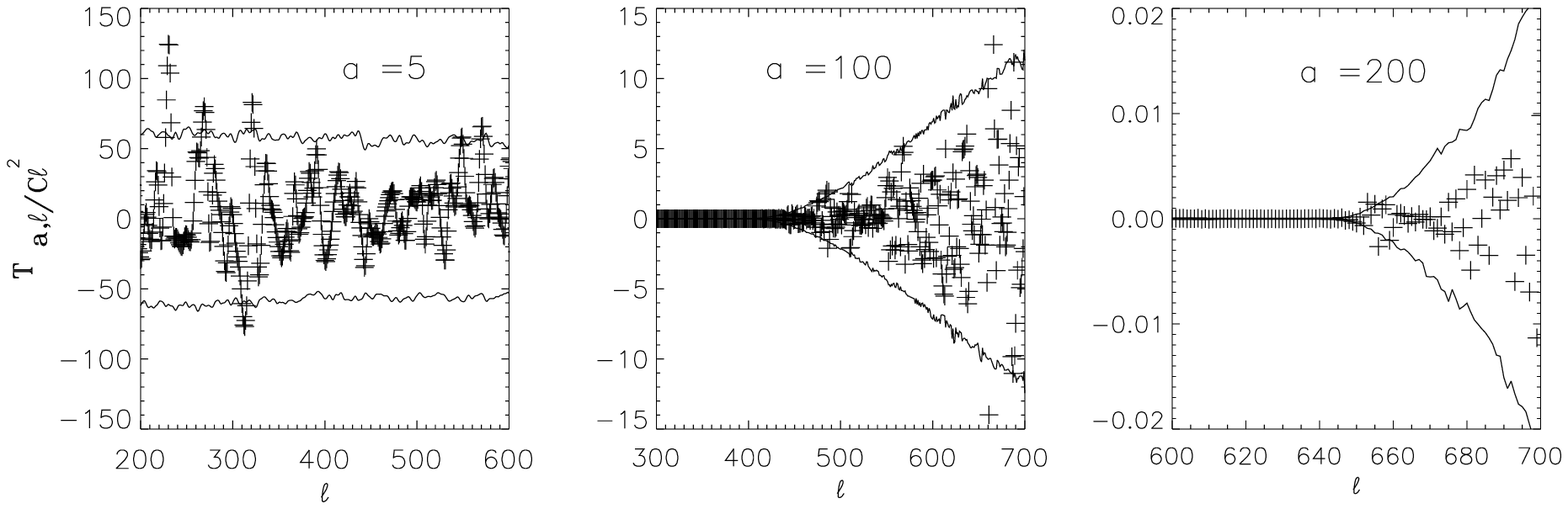,width=12cm}}
\vskip-1.4in
\centerline{\epsfig{figure=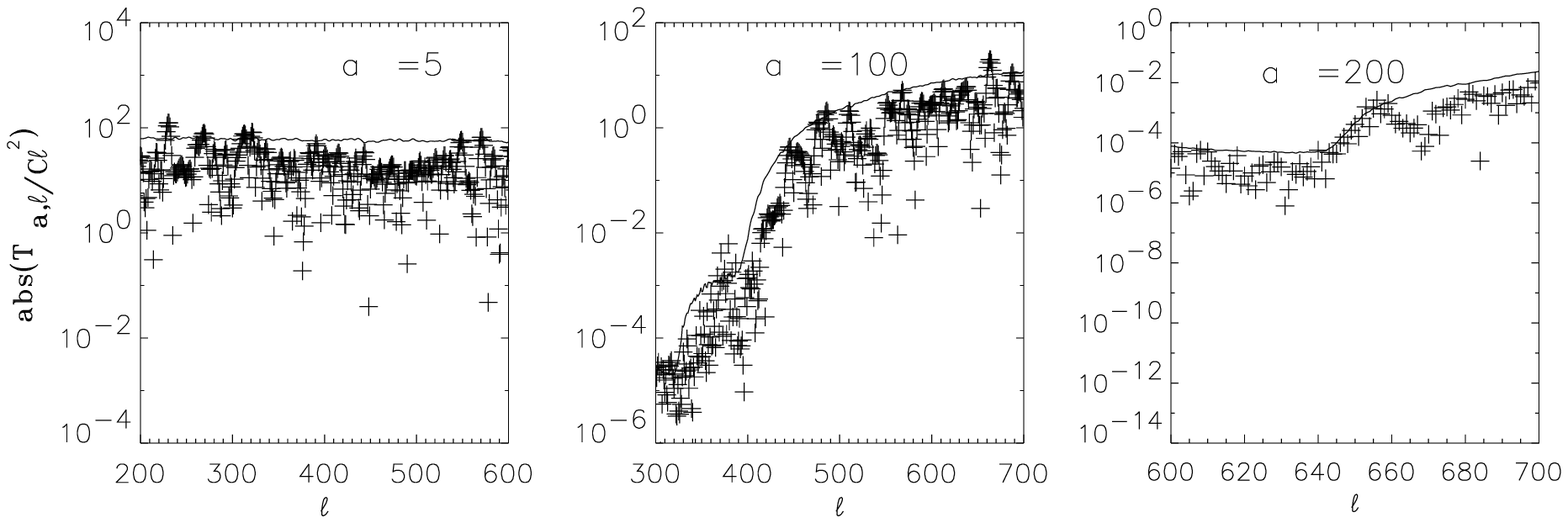,width=12cm}}
\caption{Top panel: An estimate of the orthogonalised trispectrum
corrected for Gaussian contamination and normalised ${\hat
T}_{a;\ell}/({\hat C}_\ell)^2$ for $a=5,100,200$ from the
BOOMERanG data (crosses) and the corresponding $95\%$ confidence
limits from the $1000$ Monte-Carlo simulations. Bottom panel: The
absolute value of the  estimate of the orthogonalised trispectrum
corrected for Gaussian contamination and normalised ${\hat
T}_{a;\ell}/({\hat C}_\ell)^2$ for $a=5,100,200$ from the
BOOMERanG data (crosses) and the corresponding $95\%$ confidence
limits from the $1000$ Monte-Carlo simulations. } \label{res3}
\end{figure}

Let us now proceed to the orthornomalised estimators,
${\hat T}_{a;\ell}$; a selection of estimators are plotted
for a choice of $a$s in Figure \ref{res3}. As noted above,
the $a$ are limited to $a\leq {\rm int} (l/3)$, and we see a clear
suppression of the high $a$ values for each $\ell$ (or, in the
case of the figure, of the low $\ell$ values for fixed $a$),
as the maps with limited sky coverage contain less information
than full sky maps.

\begin{figure}
\centerline{\epsfig{figure=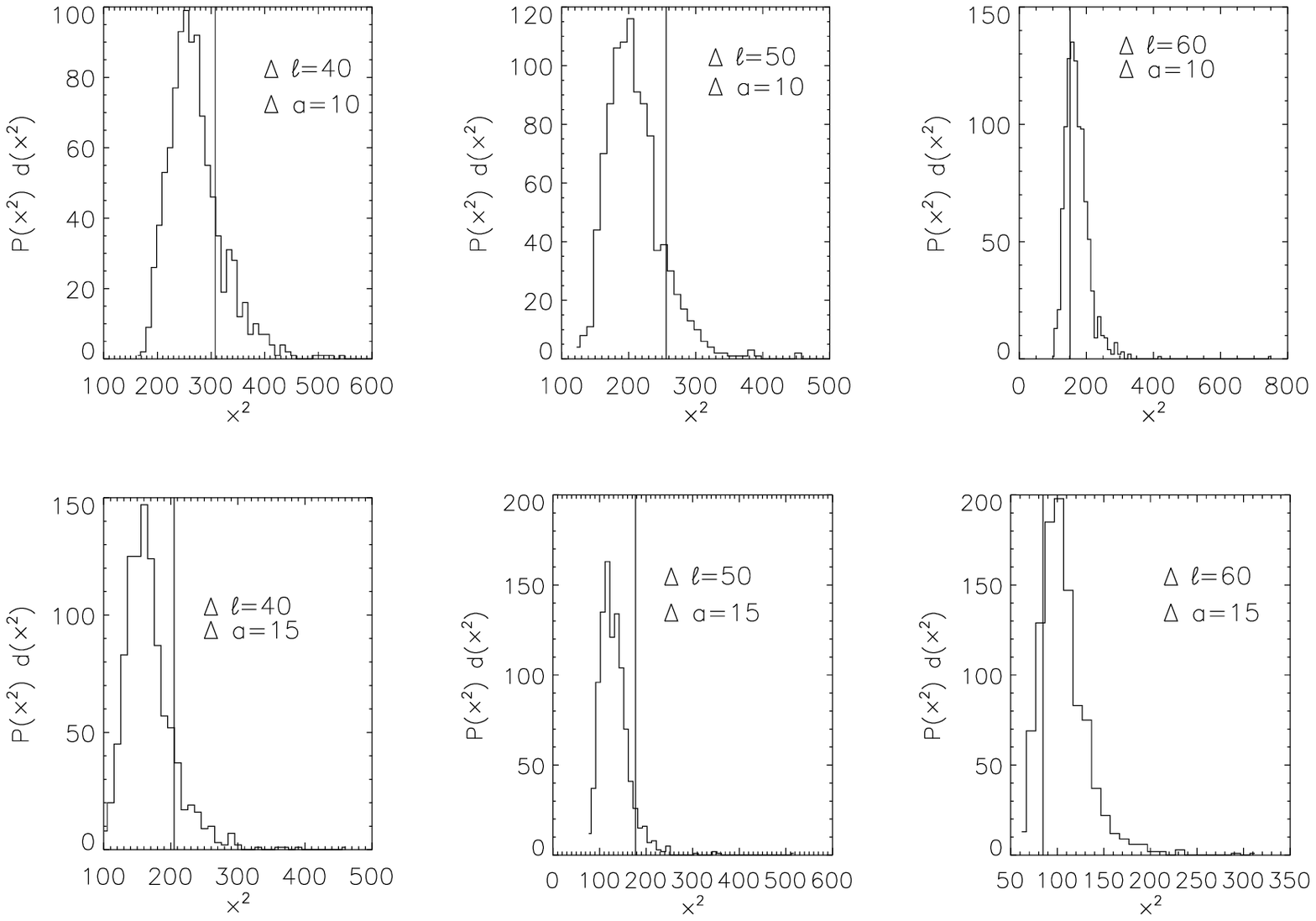,width=12cm}}
\caption{The $\chi^2$ distribution of Monte-Carlo simulated maps
(histogram) and data value (vertical line) for the trispectrum
estimator $|{\hat T}_{a;\ell}|$ in the case of $\Delta a=10$ and
$\Delta l=40,50,60$ (top) and in the case of $\Delta a=15$ and
$\Delta l=40,50,60$ (bottom)} \label{chi_abs}
\end{figure}

\begin{figure}
\centerline{\epsfig{figure=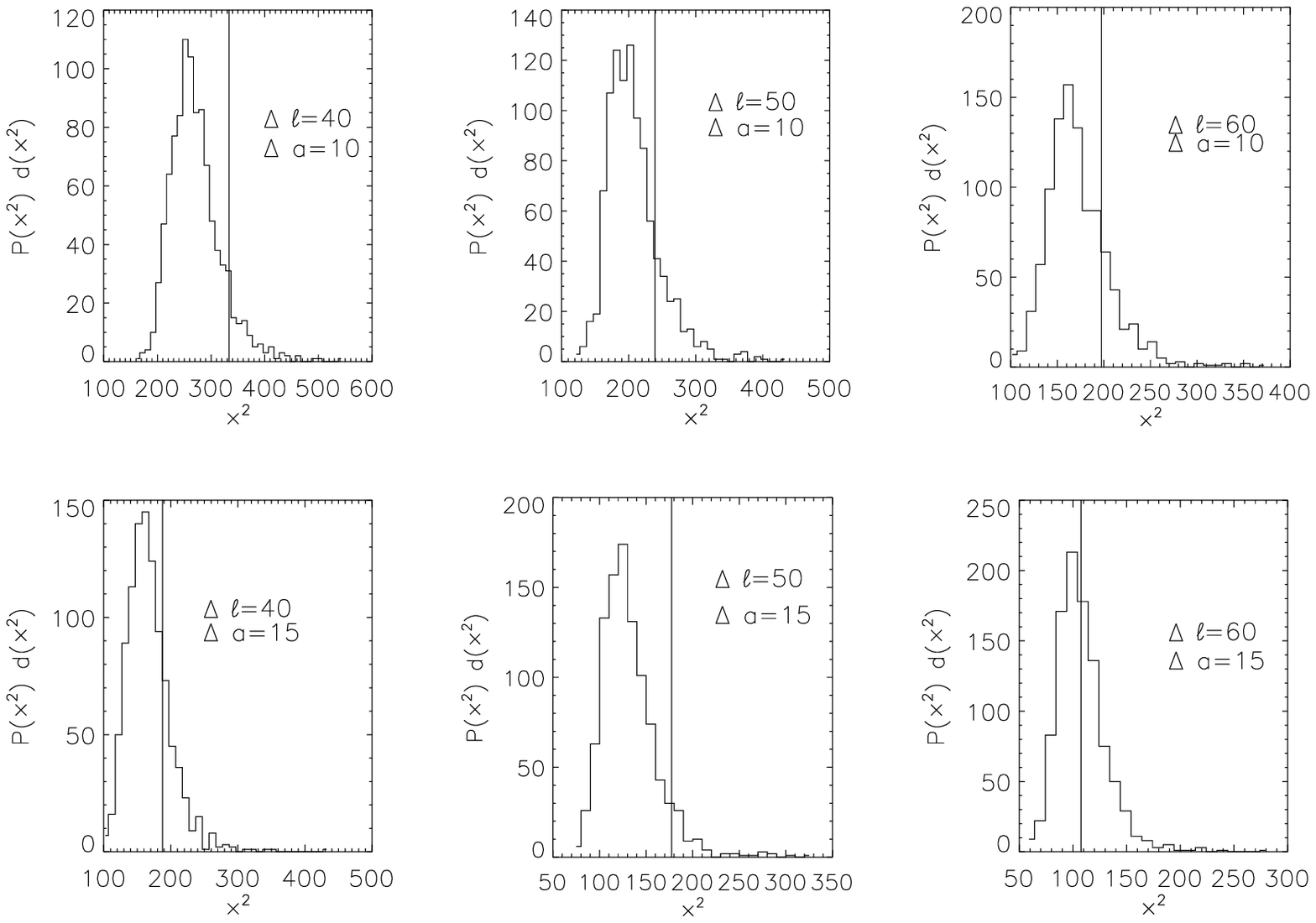,width=12cm}}
\caption{The $\chi^2$ distribution of Monte-Carlo simulated maps
(histogram) and data value (vertical line) for the trispectrum
estimator ${\hat T}_{a;\ell}$ in the case of $\Delta a=10$ and
$\Delta l=40,50,60$ (top) and in the case of $\Delta a=15$ and
$\Delta l=40,50,60$ (bottom)} \label{chi_nofs}
\end{figure}

\begin{table}
\begin{center}
\begin{tabular}[h]{|l||c|c|}
\hline
bin width &$|{\hat T}_{a;\ell}|$ &${\hat T}_{a;\ell}$  \\
\hline
 $\Delta\ell=40$ \\  $\Delta a=10 $ & $21.8\%$ &  $10.6\%$   \\
\hline
$\Delta\ell=50$ \\  $ \Delta a=10 $ & $13.7\%$ &  $19.3\%$   \\
\hline
$\Delta\ell=60$ \\  $ \Delta a=10 $ & $76\%$ &  $23.2\%$   \\
\hline
$\Delta\ell=40$ \\  $ \Delta a=15 $ & $15.7\%$ &  $26\%$   \\
\hline
$\Delta\ell=50$ \\  $ \Delta a=15 $ & $8.7\%$ & $8.5\%$    \\
\hline
$\Delta\ell=60$ \\  $ \Delta a=15 $ & $87.9\%$ & $53.8\%$  \\
\hline
%\caption{Probability that the Gaussian models have a $\chi^2$ greater
%than the data value for the both trispectrum estimators and different
%bin width}
\end{tabular}
\end{center}
\caption{Probability that the Gaussian models have a $\chi^2$ greater
than the data value for both the trispectrum estimators and for different
bin widths in $\ell$ and $a$.
\label{tab:res}}
\end{table}

Once we have calculated the trispectrum estimators both for the
BOOMERanG data and for the Monte-Carlo simulations, we can proceed to
get the $\chi^2$ distribution for the simulated Gaussian maps and compare
it to the data. We used two different approaches:
one taking as estimator ${\hat T}_{a;\ell}$ (the orthogonalised trispectrum corrected for Gaussian contamination
and normalised to $C_l^2$) and the other one taking its absolute
value $|{\hat T}_{a;\ell}|$.

We construct a standard multivariate $\chi^2$ as:
\begin{eqnarray}
\chi^2 = \sum_{\ell,\ell^{'}\!,a,a^{'}}(\langle {\hat T}_{l;a} \rangle{_G} - {\hat T}_{l;a})
C^{-1}_{\ell,a,\ell^{'}\!,a^{'}}(\langle {\hat T}_{l^{'};a^{'}} \rangle{_G} - {\hat T}_{l^{'};a^{'}})
\end{eqnarray}
deriving the expectation values $\langle {\hat T}_{l;a} \rangle{_G}$ and the
covariance matrix
$C_{\ell,a,\ell^{'}\!,a^{'}}=\langle{\hat T}_{l;a} {\hat T}_{l^{'};a^{'}}\rangle{_G}
-\langle{\hat T}_{l;a}\rangle{_G}\langle{\hat T}_{l^{'};a^{'}}\rangle{_G}$
from one of the two remaining Monte-Carlo ensembles. As discussed earlier,
we don't sum over all $\ell$ and $a$, but bin both $a$ and $\ell$, varying
the size of the bins.
Finally we calculate the $\chi^2$ distribution from the last ensemble of
Gaussian maps.

In figures \ref{chi_abs} and \ref{chi_nofs} we show the $\chi^2$
distribution derived from $1000$ Gaussian realizations compared to the 
BOOMERanG
data for both estimators and for
different bin widths. The probability $P(\chi^2>\chi_B^2)$ that a Gaussian
map has a larger $\chi^2$ than the Boomerang map is given in
table \ref{tab:res}. Although the values vary considerable with the
choice of bin-widths, none of them is below $5\%$ or $2\sigma$.
We conclude that the trispectrum does not detect any non-Gaussianity
in the coadded Boomerang 150~GHz maps.

\section{Conclusions}

We have applied an improved version of the method of \cite{cobetri} 
for measuring the trispectrum
to the four best 150~GHz Boomerang
maps. To this end, we used maps containing only one multipole each to avoid
computing the Wigner 3J symbols and subtracted the average Gaussian contribution
using an ensemble of simulated maps. We then orthogonalised these maps and normalised them to $C_\ell$.
We then binned the resulting trispectrum values with a variety
of different bin sizes, and computed the $\chi^2$ value, using a full covariance
matrix estimated from a second ensemble of simulated Gaussian maps. When
comparing the data $\chi^2$ value to the Gaussian realizations (obtained from
a third ensemble of simulated Gaussian maps) we concluded that the trispectrum
does not detect any deviations from Gaussianity.

This work complements the pixel-space analysis (Polenta et al. 2002) and the bispectrum
analysis (Contaldi et al. in preparation) of the BOOMERanG data. 

In this paper we have studied for the first time the trispectrum of
real CMB data with sub-degree resolution.
The main problem encountered was the limited sky coverage, which introduces
strong correlations, and prevents the use of orthogonalisation to remove the
Gaussian (connected) part of the trispectrum. We expect therefore that the
MAP and Planck 
satellites will be able to improve on this analysis considerably,
but this study provides a proof of feasibility for measuring the trispectrum
of full sky high resolution maps as well as first results on small angular scales.

\section*{ACKNOWLEDGEMENTS}
GDT acknowledges financial support from the Dottorato in Astronomia 
dell'Universit\`a La Sapienza and from a Marie Curie pre-doctoral
fellowship.
MK acknowledges financial support from the Swiss National Science
foundation. PGF acknowledges the support of the Royal Society.
The BOOMERanG project has been supported by Programma Nazionale di
Ricerche in Antartide, Universit\'a di Roma ``La Sapienza'', and
Agenzia Spaziale Italiana in Italy, by NASA and by NSF OPP in the
U.S. We acknowledge the use of the HEALPix package and of the Oxford 
Beowulf cluster for our computations. We thank Andrew Jaffe and Alessandro
Melchiorri for their helpful comments and Jonathan Patterson for his
help.

\end{document}